# Fast Mining of Spatial Frequent Wordset from Social Databases

Yongmi Lee[1]    Kwang Woo Nam[2]    Keun Ho Ryu[1]

Abstract: In this paper, we propose an algorithm that extracts spatial frequent patterns to explain the relative characteristics of a specific location from the available social data. This paper proposes a spatial social data model which includes spatial social data, spatial support, spatial frequent patterns, spatial partition, and spatial clustering; these concepts are used for describing the exploration algorithm of spatial frequent patterns. With these defined concepts as the foundation, an SFP-tree structure that maintains not only the frequent words but also the frequent cells was proposed, and an SFP-growth algorithm that explores the frequent patterns on the basis of this SFP-tree was proposed.

## 1. Introduction

The development of Internet and mobile technologies has led to the production of enormous amount of information along with increased interest in the social network service. In typical micro blogs such as Twitter and Facebook, users form social networks by sharing information consisting of short text unlike previous web documents [1, 2]. Since such social data is dynamic in comparison with web documents and most data includes spatial information including GPS locations, it has a characteristic of transaction data. However, such social data mostly consists of text, so only a simple attempt such as change in the frequent count has been made in the studies applying the previous data exploration algorithms [3, 4, 5, 6, 7].

Social data in micro blogs contains much shorter but semantically implied information than web documents. In other words, the average number of words per tweet in Twitter is 14.98 words and each tweet consists of 1.4 sentences on average [8]. This indicates that the association rule exploration algorithms in the previous data mining field can be easily applied to spatial social database due to the characteristics of micro blog information. Such fact means that a spatial frequent wordset can be found from social database by expanding the association rule algorithm in data mining field to support space. Generally, it is not easy to infer a meaning beyond intuitive facts such as "~ is frequent' from information explored as a frequent pattern. However, if spatial information can be combined, it could be more precise knowledge. In other words, even if a certain frequent pattern was a noticeably frequent pattern in the whole database, it may not be a frequent pattern in a certain space. This is possible because a significant amount of social data include spatial information which can be mapped as position. If word patterns used together frequently in a specific space can be explored from social data, regional characteristics, such as relative political, social interest or trend in a specific space, can be easily identified.

Corresponding author :

    Kwang Woo Nam    kwnam@kunsan.ac.kr

1 Department of Software, Chungbuk National University, Cheongju, South Korea

2 Department of Computer and Information Engineering, Kunsan National University, Gunsan, South Korea

Figure 1 shows micro blog data in the real world (a) and an example (b) of spatial social data stored in the database. For convenience, let's substitute administrative districts (a~f) in the real world where social data is created for spatial information such as the GPS location of the data. For example, social data of which object ID was 'o2' was created in the administrative district 'b'. Such data is saved in the form of transaction database through preprocessing process as shown in Figure 1(b). For example, each tuple of spatial social database can include object ID, wordset and spatial information.

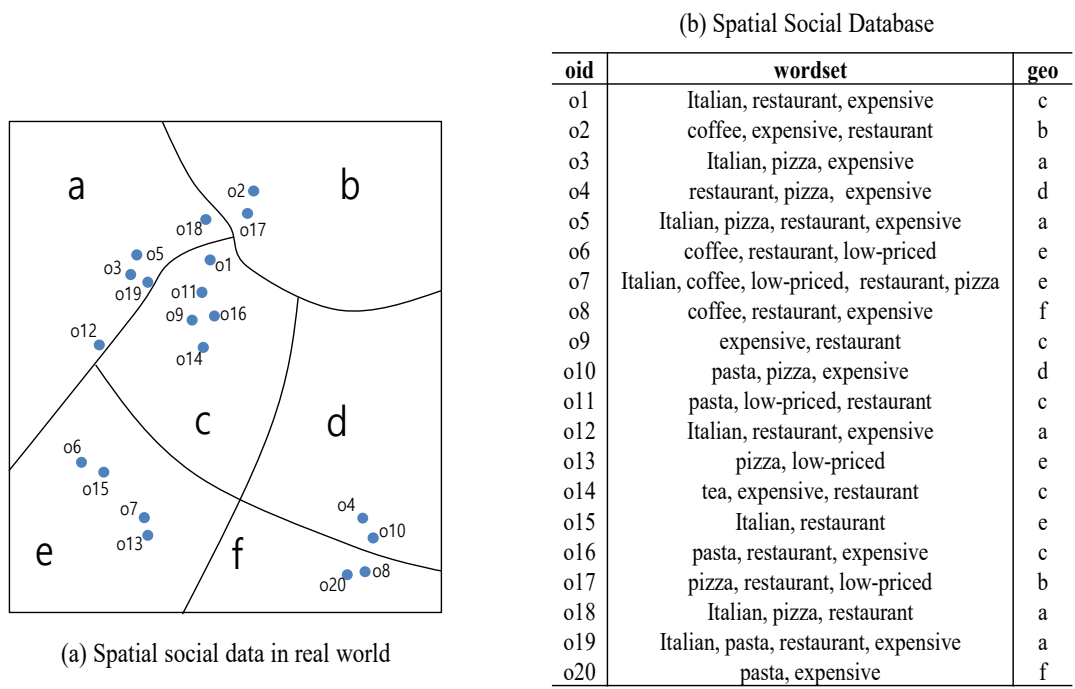

(b) Spatial Social Database

| oid | wordset | geo |
|---|---|---|
| o1 | Italian, restaurant, expensive | c |
| o2 | coffee, expensive, restaurant | b |
| o3 | Italian, pizza, expensive | a |
| o4 | restaurant, pizza, expensive | d |
| o5 | Italian, pizza, restaurant, expensive | a |
| o6 | coffee, restaurant, low-priced | e |
| o7 | Italian, coffee, low-priced, restaurant, pizza | e |
| o8 | coffee, restaurant, expensive | f |
| o9 | expensive, restaurant | c |
| o10 | pasta, pizza, expensive | d |
| o11 | pasta, low-priced, restaurant | c |
| o12 | Italian, restaurant, expensive | a |
| o13 | pizza, low-priced | e |
| o14 | tea, expensive, restaurant | c |
| o15 | Italian, restaurant | e |
| o16 | pasta, restaurant, expensive | c |
| o17 | pizza, restaurant, low-priced | b |
| o18 | Italian, pizza, restaurant | a |
| o19 | Italian, pasta, restaurant, expensive | a |
| o20 | pasta, expensive | f |

(a) Spatial social data in real world

**Figure 1 Social data in real world and a social database**

Figure 2 shows examples of pattern that can be explored from spatial social database in Figure 1 when the minimal support threshold for becoming a frequent pattern is 2. The previous frequent pattern algorithm does not consider spatial attributes. Therefore, only the information regarding non-spatial frequent pattern can be found as shown in Figure 2(a). In other words, the word which is used most frequently in the entire database is {restaurant}, and we can see intuitively the fact that the wordset {restaurant, expensive} is more frequent than the wordset {restaurant, low-priced}. However, it is not easy to infer the knowledge regarding the spatial frequent pattern inherent in social database. Let's examine the frequent patterns in Figure 2(b), the spatial frequent wordset, in this study. We can see that 'wordset {restaurant, expensive} from the entries of ({restaurant, expensive}, c, 4) and ({restaurant, expensive}, a, 3) in this example is only frequent in Areas c and a. As a result, the wordset {restaurant, expensive}, which appeared to be frequent in Figure 1.2(a), was frequent only in Areas c and a in Figure 1.2(b). In other words, we can assume from this spatial frequent wordset that the restaurants in Areas c and a are expensive.

The purpose of this study is to find an algorithm which explores a spatial frequent pattern that explains relevant characteristics of specific space from spatial social data as shown in the example above. However, according to the precedent studies, almost no study has been carried out with regard to the spatial frequent pattern. Spatial web object studies regarding documents on the file level that included spatial information on the Internet focused on search using index or spatial keyword for efficient spatial query. Studies regarding spatial association rule and co-location discussed studies regarding the frequent pattern of phase relationship between spatial objects but such studies targeted the entire space, not specific space. In addition, studies to detect items such as trends of social interests and regional accidents such as a natural disaster from social data are being carried out, but these studies are merely studies regarding the application of the existing algorithms. Apriori [9] or FP-growth [10] algorithms in traditional data mining also handle frequent patterns but spatial frequent patterns that can explain the characteristics of a specific space are not covered. Therefore, mining of spatial frequent pattern is a new research field presented in this study [11].

(a) Simple frequent patterns

| word | count |
|---|---|
| restaurant | 16 |
| expensive | 13 |
| Italian | 8 |
| pizza | 8 |
| low-priced | 5 |
| pasta | 5 |
| coffee | 4 |
| restaurant, expensive | 10 |
| restaurant, Italian | 7 |
| restaurant, pizza | 5 |
| restaurant, low-priced | 4 |
| restaurant, coffee | 4 |
| Italian, expensive | 5 |
| pasta, expensive | 4 |
| restaurant, pasta | 3 |
| pizza, Italian | 4 |
| pizza, expensive | 4 |
| pizza, low-priced | 3 |
| : | : |

(b) Frequent patterns on the real world

| word | geo | count |
|---|---|---|
| restaurant | c | 5 |
| restaurant | a | 4 |
| restaurant | e | 3 |
| restaurant | b | 2 |
| expensive | a | 4 |
| expensive | c | 4 |
| expensive | d | 2 |
| expensive | f | 2 |
| Italian | a | 5 |
| Italian | e | 2 |
| pizza | a | 3 |
| : | : | : |
| restaurant, expensive | c | 4 |
| restaurant, expensive | a | 3 |
| restaurant, Italian | a | 4 |
| restaurant, Italian | e | 2 |
| restaurant, pizza | a | 2 |
| restaurant, low-priced | e | 2 |
| : | : | : |

**Figure 2. Frequent Wordset and Spatial Frequent Wordset**

In order to solve this, FP-tree and FP-growth algorithms that were relatively easy to be expanded to include frequent spatial information, among traditional frequent pattern exploration algorithms, were expanded in this study. For a strategy to include information regarding frequent space in the algorithm just as frequent words in FP-tree, a cell of which entire space was divided by grid was defined as basic unit to simplify the calculation. Also, SFP-tree where frequent words and frequent spaces are saved was constructed based on such strategy and the SFP-growth algorithm which explored a frequent pattern based on such tree was proposed.

## 2. Related Work

A frequent pattern is a set of items which have the frequency over predefined minimal support threshold [10]. This forms base knowledge for mining an association rule or a sequential pattern from the transaction database consisting of merchandise trade records.

| id | wordset |
|---|---|
| o1 | Italian, restaurant, expensive |
| o2 | coffee, expensive, restaurant |
| o3 | Italian, pizza, expensive |
| o4 | restaurant, pizza, expensive |
| o5 | Italian, restaurant |

**Figure 3. An Example of Social Database**

For example, when traditional association rule mining is applied to normal social database as shown in Figure 3, frequent patterns such as '{restaurant, expensive} : 4' are explored in such process and an association rule is created based on such patterns. The best-known algorithm mining frequent patterns is the Apriori [8] algorithm. Apriori algorithm creates a frequent pattern candidate set for all levels and checks the frequency of such candidate set through database scanning repeatedly. Algorithms such as DHP [12], Eclat [13] and DIC [14] were presented in order to reduce overheads which occurred in this process. DHP minimized overheads regarding the creation of candidate sequence using a hash-based algorithm, and DIC carries out the creation of a candidate set and a frequent pattern at the same time to minimize the repeat count.

Generally, Apriori algorithms accompany repetitive database scanning. Contrarily, FP-tree-based FP-growth [9] algorithm minimizes multiple DB scanning, establishes frequent prefix tree by scanning DB twice and explores a frequent pattern based on this tree. This algorithm calculates frequent items through the first database scanning and configures FP-tree consisting of only frequent items through the second database scanning.

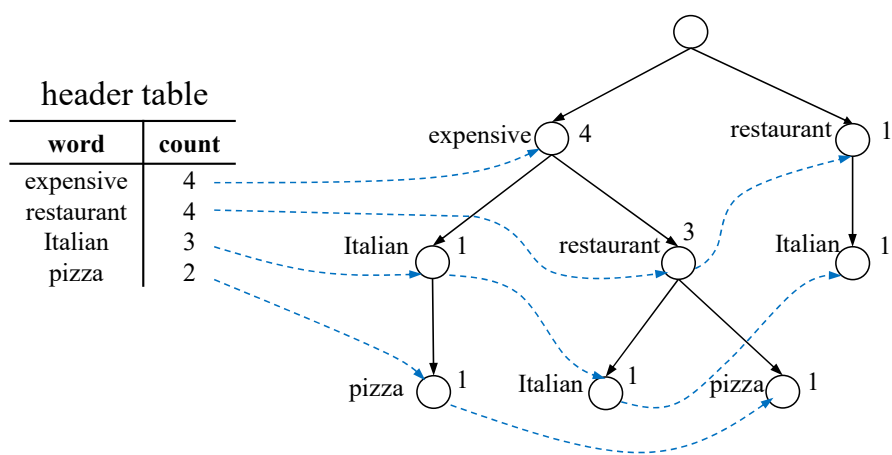

**Figure 4. Construction of FP-tree from Social Database**

Figure 4 shows an example of establishing FP-tree using the social database in Figure 3. FP-tree is constructed as a tree establishing the header table consisting of tables and wordsets in the database in the form of prefix tree by calculating the support count at a word level. At this time, each word entry in the header table has a link which connects the relevant word nodes existing in the pattern tree sequentially. Also, the FP-growth algorithm uses the divide and conquer strategy to explore a frequent pattern from FP-tree, and it establishes conditional FP-tree from the word with the smallest support count to the word with the largest support count and explores a frequent pattern recursively.

As a result, FP-growth algorithm showed better time performance than Apriori algorithms, so this algorithm was expanded in wide areas in many studies. For example, algorithms such as FreeSpan

[15] or PrefixSpan [16] in the sequential pattern mining and FP-stream [17] or SWIM [18] in a study regarding the frequent pattern in data stream area are being studied in various aspects. The frequent pattern which this algorithm explores is the type of {w1, w2, ..., wn}. This is similar to social data where the research area of this algorithm has an aspect of transaction data, but it means that a specific space cannot be reflected. In other words, a frequent pattern explored by FP-growth algorithm is a global pattern which cannot explain the characteristics of a specific space. However, this study focuses on the FP-growth algorithm as the spatial frequent pattern-based algorithm for two reasons. First, this algorithm shows generally good time performance. Second, the structure of FP-tree constructed in the form of header table and prefix tree is appropriate to be expanded to an algorithm which includes frequent spatial information.

## 3. Problem Definition

When the message attribute of micro blog "text" is a wordset consisting of n/0 words, spatial social data "ssd" where spatial information "geo" of social data, such as GPS, is tagged is a tuple constructed as follows.

$$ssd := (\ oid,\ wordset:\{\ w_1,\ w_2,\ ...,\ w_n\ \},\ geo\ )$$

The spatial information of social data created from social network is expressed in geo: $(x, y)$ which is a pair of longitude and latitude in double type. For example, if micro blog data such as ($o_{15}$, 'we are in an Italian restaurant', p15) exists, it is constructed as a group of words after text parsing process and removal of stopword which has no significance in the meaning. $o_{15}$ is converted into ($o_{15}$, {Italian, Restaurant}, p15).

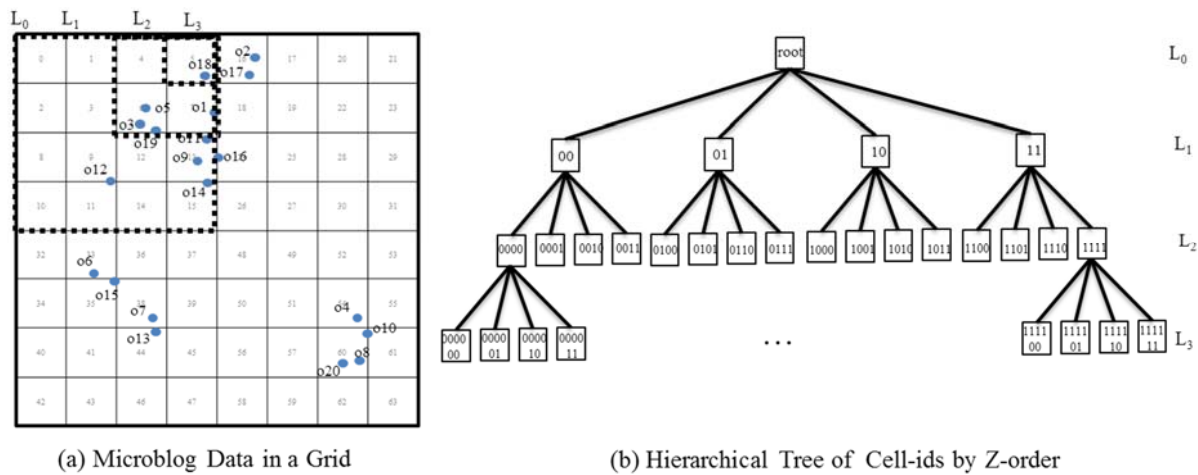

(a) Microblog Data in a Grid    (b) Hierarchical Tree of Cell-ids by Z-order

**Figure 5. An Example of Spatial Partitioning and Cell-id Hierarchy**

For efficient mining process, the entire space is divided into cells in a 2D grid using the space filling curve as shown in Figure 5(a) and these cells are grouped as cells where each micro blog data is included. At this time, *gid,* which is an integer type value, is assigned to each spatial cell by z-ordering. This *gid* divides one side into 4 and each divided side has hierarchical characteristics as shown in Figure 5(b) according to the characteristics of z-order which can allocates a bit value including 00, 01, 10 and 11 recursively. For example, $o_3$ and $o_5$ are positioned in No. 6 cell of $L_3$ level of which *gid* is 000110 as shown in Figure 6(a). Therefore, *($o_3$, {Italian, pizza, expensive}, $p_3$) and ($o_5$,*

*Italian, pizza, expensive, restaurant}, $p_5$)* are spatially grouped into *($o_3$, {Italian, pizza, expensive}, $gid_{000110}$)* and *($o_5$, Italian, pizza, expensive, restaurant}, $gid_{000110}$)* that have the same $gid_{000110}$. In addition, when $gid_{000110}$ is allocated according to the spatial hierarchical characteristics of gid, it can be easily identified that it is located at 00 in $L_1$, 01 in $L_2$ and 10 in $L_3$. Micro blog data can be spatially grouped very efficiently by setting the spatial partition unit of the lowest level in the smallest size that can classify a meaningful wordset using these characteristics and merging to higher hierarchy in bottom-up form. At this time, various sizes can be selected for the lowest level cell according to the purpose of data or analysis. The most common method is to set a meaningful distance (for example, 100m, 200m, 1km) according to the distance in the real world and determine a size where the space is divided in the minimum unit that is close to the distance. Another method is to determine the average number of micro blog data per cell (for example, 100 and 1,000) and decide the size of divided space based on the number. For the convenience in understanding, it is explained in this study that it is determined based on the distance.

The support count *sc* is how many times a specific wordset appears in non-spatial association rule mining regardless of space, and if *sc>c,* it is called frequent itemset and c connotes the minimum support count. Here, the user sets *c* depending on whether a wordset which appears over a certain number of times is considered as meaningful or not.

Let's expand this concept into a spatial wordset. A spatially meaningful wordset in micro blog social data appears repeatedly in proximate space. Therefore, let's set the minimum unit of proximate space as the cell size at the lowest level in the spatial partition grid and define the spatial support count *ssc* regarding how many times it appears in this cell. And let's define *ssc* > $\sigma$ in a certain wordset as spatial frequent wordset and define $\sigma$ as minimum spatial support count. For example, we can see that *ssc* of {Italian} {expensive} and {Italian, expensive} in No. 6 cell is 3 according to $o_3$, $o_5$, and $o_{19}$, and *ssc* of {pizza} and {restaurant} is 2. When the minimum spatial support count is $\sigma$ = 3 to find a spatial frequent wordset, {pizza} and {restaurant} of which ssc is less than 2 are discarded and only {Italian} {expensive} and {Italian, expensive} of which ssc is over 3 is a meaningful spatial frequent wordset.

When a certain spatial wordset $sfws_{Li}$ exists in the specific *ith* level hierarchy, $L_i$, in spatial grouping hierarchy as shown in Figure 5(b), the spatial frequent wordset mining in this study finds the spatial frequent wordset which is $|sfws_{Li}|$ > $\sigma_{Li}$ on spatial social database *SSDB*. Here, $\sigma_{Li}$ means the minimum spatial support count on the spatial hierarchy $L_i$. The next paragraph explains the spatial frequent wordset mining method based on SFP-tree.

## 4. Mining of Spatial Frequent Wordset

### 4.1 SFP-tree

SFP-tree consists of spatial prefix tree, where the information regarding the spatial frequent wordset in each cell and the appearance count are saved, and the header table, which contains the information regarding 1word-spatial cell frequency. It is the same with FP-tree in that all parent nodes of a specific node consist of common prefix words of each node. However, it is distinguished in that each node has space cells, where such word appears frequently, and an auxiliary table for the appearance count at such cell. SFP-tree can be defined as follows.

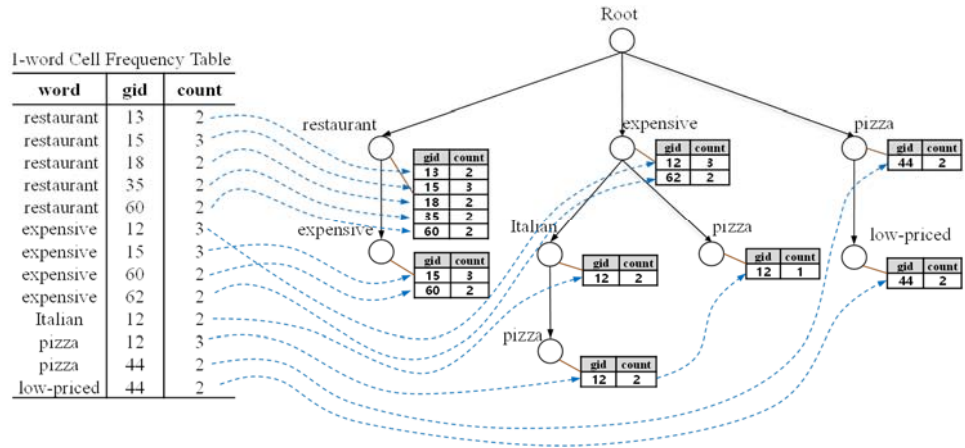

Figure 6. A Partial SFP-tree for Example Databases

1. SFP-tree consists of prefix tree *sfptree* which consists of nodes that contain information of spatial frequent words and $FT_{1\text{-}word\text{-}cell}$ which is the header table of spatial frequent word nodes.

2. When the information of word is *w*, the link for parent node is *uplink*, the set of links for child nodes is *downlinks* and information table of space cell where word *w* in each node appears frequently and the appearance frequency in such cell is *gidCountTable*, each node of *sfptree* is as follows.

    *(w, uplink, downlinks, gidCountTable )*

3. Here, *gidCountTable* is a set of *(gid, count)* which is a pair of *gid*, which is the space cell identifier where word *w* appears frequently, and *count*, which is the appearance frequency at such cell.

4. Child nodes are prefix trees that share the words of parent node commonly and the word *w* of the top level node is *null*.

5. When the total sum of frequency where the spatial frequent word *w* appears at the space cell *gid* in entire SFP-tree is *count* and the set of links for all nodes that include this spatial frequent word is *nodelinks*, $FT_{1\text{-}word\text{-}cell}$ is the set of *(w, gid, count, nodelinks)*.

### 4.2 Construction of SFP-tree

Algorithm 1 shows *mineSFWordset* algorithm for establishing SFP-tree and finding a spatial frequent wordset using SFP-tree. *mineSFWordset* takes the input of spatial social database *SSDB*, the minimum spatial support count $\sigma$ and the spatial partition hierarchical type grid *grid*. This algorithm creates an empty *sfpTree* first (1) and constructs the global frequent word table $FT_{1\text{-}word}$ and the spatial frequent word table $FT_{1\text{-}word\text{-}cell}$ which consists of only words that are frequent in one space cell over $\sigma$ through the entire spatial social database scanning using *constructFTs* (2). Next, this algorithm traverses the entire database again based on these two frequent tables and constructs SFP-tree consisting of only frequent words and spaces using each *ssd*. At first, text in each social data is converted into the set of words *wordset* through the root extraction and stopword removal process from the words. Also, a lot of costs are required for processing each word as a string.

Therefore, *wordset* is converted into a set of wid values that are identifiers on the word dictionary.

**Algorithm 1 :** *mineSFWordset( SSDB, σ, grid )*
**Input :**   Spatial Social Database, SSDB
           Spatial Minimum Support Count, σ
           Hierarchy Tree of Spatial Grid, grid
**Output :** *A Set of Spatial Frequent Wordset, SFW*
  1: *sfpTree* ← create a New and Blank SFP-tree;
  2: *(FT$_{1\text{-}word}$, FT$_{1\text{-}word\text{-}cell}$* ← *constructFTs(SSDB, σ);*
  3: **for each** *ssd* ∈ **SSDB do**
  4:   *wids* ← *refineAndSort( ssd.wordset, FT$_{1\text{-}word}$ );*
  5:   *gid* ← *spatialGrouping( grid, ssd.geo );*
  6:   *insertSSD( sfpTree, FT$_{1\text{-}word\text{-}cell}$, wids, gid );*
  7: **end for**
  8: *SFW* ← *SFP-Growth( sfpTree, FT$_{1\text{-}word\text{-}cell}$ , σ, grid);*
  9: **return** *SFW*;

At this time, only words in $FT_{1\text{-}word}$ are selected in order to use only words that appear a meaningful number of times in the relevant social database. Also, it is necessary to reduce the total number of nodes in SFP-tree for efficient spatial frequent wordset mining. For such purpose, words are sorted in order from a word with larger frequent count to a word with smaller frequent count so that words that appear frequently in the entire area become patient nodes and words that appear less frequently in the entire area become child nodes, and then words are converted to *wids* (3). The spatial coordinates of *ssd* are mapped in spatial partition unit on the hierarchical spatial grid *grid* and *gid* is allocated (5). A new wordset data is inserted by entering *wids* and *gid* of *ssd* on *sfpTree* (6). If the traversal for all data in the social database is finished, the construction of *sfpTree* is completed. Mining is finished when spatial wordset is extracted and returned, using *sfpTree* and SFP-growth (8).

```
Algorithm 2 : InsertSSD( sfpTree, FT_{1-word-cell}, wids, gid )
Input :   Root node of SFPTree, root;
          1-word Cell Frequency Table of SFPTree, FT_{1-word-cell};
       A set of refined and sorted-by-count words(wordset), wids;
       Gid of the wordset, gid;
Output : none
 1: parentNode ← sfpTree.root;
 2: for each w ∈ wids do
 3:   node ← parentNode.findChildNodes( w );
 4:   if ( node == null ) then
 5:     gidCountTable ← new hash map;
 6:     gidCountTable.put( gid, 1 );
 7:     node ← makeNewNode(w, gidCountTable )
 8:     parentNode.add( node );
 9:   else
10:     gidCountTable ← node.getGidCountTable();
11:     count ← gidCountTable.get( gid );
12:     if ( count == null ) then
13:       gidCountTable.put( gid, 1 );
14:     else
15:       gidCountTable.put( gid, count+1 );
16:     end if
17:   end if
18:   FT_{1-word-cell}.addLink( w, gid, node );
19:   parentNode ← node;
20: end for
21: return;
```

The Insertion of *sfpTree* takes the input of *root* of *sfpTree* and $FT_{1\text{-}word\text{-}cell}$, and *wids* and *gid* of *ssd* you wish to insert into *sfpTree*. The insertion of *w* which is *w*∈*wids* is classified into three types. First, if a child node which corresponds to a specific word *w* does not exist among wids you wish to insert (4), create *gidCountTable* (5), insert *(gid, 1)* entry (6), create a new node which has *gidCounTable* (7) and add it as a child node to the patient node (8). Second, if *gid* entry does not exist in *gidCountTable* even if a child node which corresponds to *w* exists (11-12), add an entry to save *(gid, 1)* (13). Third, if *gid* entry exists in both the child node which corresponds to the word and *gidCountTable,* increase the support count of *gid* entry in *gidCountTable* by 1 (15). Lastly, when the *(w, gid)* entry and the relevant node in $FT_{1\text{-}word\text{-}cell}$ are connected with a pointer (18-19), the insertion algorithm is finished.

### 4.3 SFP-Growth

*SFP-Growth* is a pattern exploration algorithm which finds all spatial frequent wordsets of which frequent count is over σ in the spatial hierarchical tree from *sfpTree* and $FT_{1\text{-}word\text{-}cell}$. In other words, when a spatial hierarchical tree of which height is *h* exists as shown in Figure 5(b), this algorithm explores all spatial frequent wordsets from the lowest level *h-1* to a higher level *0*. $FT_{1\text{-}word\text{-}cell}$ contains the information regarding frequent words and frequent count in the cell-unit space of the lowest hierarchy and links for all nodes in *sfpTree* where each 1-cell frequent word *(w, gid)* of *(h-1)* hierarchy is included. *SFP-Growth* extracts spatial word conditional FP-tree of 1-cell frequent word *(w, gid)* in each hierarchy using the link of nodes and performs the *FP-Growth* algorithm based on the extracted spatial word conditional FP-tree, mining a spatial wordset which is used with word *w* in the spatial cell *gid*.

**Algorithm 3:** SFP-Growth( sfpTree, FT$_{1\text{-word-cell}}$, σ, grid )
**Input :** a SFP-Tree, sfpTree;
        Minimum spatial support countl, σ;
**Output :** Spatial Frequent Wordset, SFW
  1: SFWordset ← ∅
  2: i ← grid.height - 1;
  3: **for** i ≥ 0 **do**
  4:   FT$_{Li}$ ← spatialGeneralization(FT$_{1\text{-word-cell}}$, grid, i);
  5:   FT$_{reverse}$ ← iterateReverse( FT$_{Li}$);
  6:   **for each** ( (wid, gid), count ) ∈ FT$_{reverse}$ **do**
  7:     (condTree, FT$_{condTree}$) ← spatialWordCondTree(sfpTree, FT$_{Li}$, wid, gid, σ );
  8:     SFWordset ← SFWordset ∪ FP-Growth(condTree, FT$_{condTree}$);
12: **end for**
13: i--;
14: **end for**
15: **return** SFWordset;

Algorithm 3 shows the *SFP-Growth* function which explores *SFWordsets* that are the group of spatial frequent wordsets based on *sfpTree*, *FT$_{1\text{-word-cell}}$*, and σ. This algorithm repeatedly traverses from the lowest hierarchical cells to the top level hierarchical cell, in a bottom-up manner, *h* times, at *grid* hierarchical tree of which height is *h*, and explores a spatial frequent wordset for each *i* hierarchy (2-3). 1-word cell table *FT$_{Li}$* is created in each *i* hierarchy using *FT$_{1\text{-word-cell}}$* that is the frequent word table in the lowest hierarchy through *spatialGeneralization* in each *i* hierarchy. *FT$_{Li}$* is a set of *(w, gid, count)* which is a tuple of *w* and *gid*, the frequent words in i hierarchy, and *count*, the frequent count. Next, header table *FT$_{condTree}$* of spatial word condition tree *condTree and condTree* is extracted from all spatial words *(w, gid)* in *FT$_{Li}$* using *spatialWordCondTree*. In order to minimize the size of *condTree* created in each traversal, inverse traversal is carried out in order from a word with smaller *count* to a word with larger count (5-6). The spatial word condition tree is non-spatial prefix tree that only extracts the tree part which satisfies a specific word *w* and a specific space *gid* from *sfpTree* which includes spatial information. Based on this, a spatial wordset which is used with the word *w* together is explored from the space cell *gid* by executing the *FP-Growth* algorithm.

## 5. Performance Evaluation

### 5.1 Social Data and Environments

Data used in the experiments was collected directly through Twitter Streaming API. When a specific condition is specified, Twitter Streaming API continuously transmits tweets that satisfy the relevant condition to the client as stream. The transmitted data may be a part of tweets which satisfy the relevant condition, not all tweets. It is announced that if the amount of data is too great, Twitter selects and transmits data randomly.

Table 1 states the scope of target social data for experiments. The collected data is Twitter data generated in North America including U.S. and Canada generated from November 18, 2012 to November 30, 2012. It is 11 days and 10 hours (274 hours) in time. The spatial boundary of this data is from (-51.36, 78.58) to (-141.89, 14.55) that includes the entire North America. The number of social transactions is approximately 10 million transactions and 39,000 transactions per hour. The number of words (unique words) in the word dictionary is approximately 4.2 million words that exceed the number of words that are used normally. It's because userid, url, numbers and foreign languages are included in the text of social data in addition to common words. The size of DB file on the hard disk is approximately 43.9 GB.

Table 1 Social Data for Experiments

| Standard | Scope |
|---|---|
| Spatial scope | (-51.36, 78.58) ~ (-141.89, 14.55) |
| Temporal scope | 2012. 11. 18 17:15:22 - 2012. 11. 30 03:11:38 (approximately 274 hours) |
| Number of data | 10,713,671(39,101/hour) |
| Type of word | 4,235,664 unique words |

In this experiment, the Windows 7 64 bit system with Intel i7 3770 CPU and 16 GB memory was used and normal 7200 RPM HDD was used. Mongo DB 2.2.1 version was used for saving social data and it was implemented using Java 1.7 version. And, Apache Lucene Version 4 was used for the text analyzer.

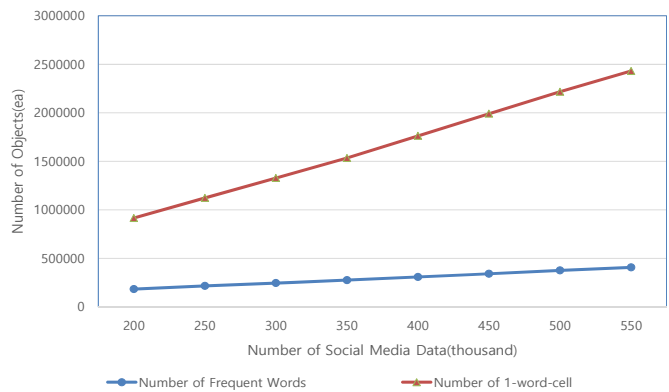

**Figure 7. Number of Frequent Wordset and 1-word-cell**

### 5.2 Effect of Data Size

The graph in Figure 7 shows the 1st frequent data change as the number of social data increases. The number of frequent words means the number of entries in each table. As shown in the graph, the number of frequent words and the number of frequent words per cell increase linearly as the number of social data increases. At this time, the number of frequent words per cell increases faster than the number of frequent words because the number of frequent words per cell increases continuously by "average number of cells * number of frequent words," and this is the expected result.

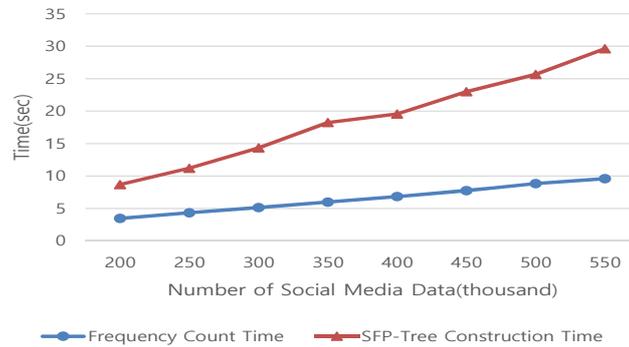

**Figure 8. Counting and SFP-tree Construction Time**

Figure 8 shows how the time to create frequent tables for data and SFP-tree is affected as the number of social data increases. It is exactly proportional (1:1) to the number of data which constructs the frequent table through the first DB scanning before SFP-tree is constructed. In comparison, the SFP-tree construction time increases in proportion to the number of data through the second DB scanning, but the slope is much steeper than the frequent table construction. For example, the frequent count time was less than 5 seconds when the number of data was 250,000, and it is close to 10 seconds which is almost doubled when the number of data is 500,000. On the other hand, the tree construction time is 10 seconds when the number of data is 250,000, and it exceeds 25 seconds when the number of data is 500,000. This is because the construction of frequent table only includes DB scanning time, but the SFP-tree construction time includes DB scanning time and the time to search and insert a node of SFP-tree.

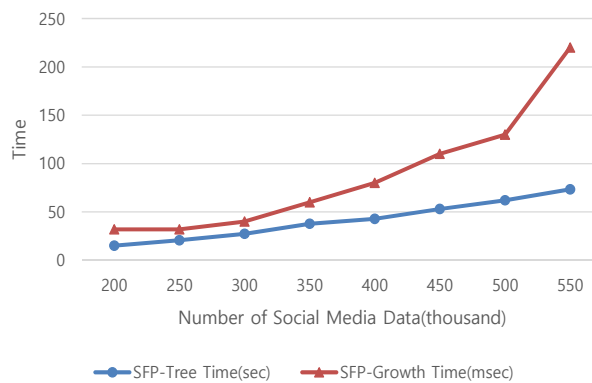

**Figure 9. SFP-Tree Construction and SFP-Growth Time**

Figure 9 shows the data construction time (+time to discover a dense cluster) according to the number of social data and the time of frequent pattern inference using actual SFP-growth. Here, the SFP-tree construction time is in second and the SFP-growth time is in msec. Graphs in different units were used in order to confirm the trend. In these graphs, the tree construction time increases linearly in proportion to the DB scanning time as the number of data increases. On the other hand, SFP-growth shows a similar rate of increase with $O(N\log N)$ in the SFP-growth time because this includes repetitive recursive arithmetic operation for the tree.

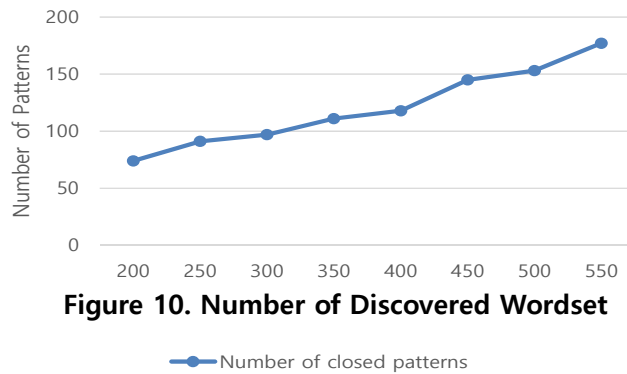

**Figure 10. Number of Discovered Wordset**

Figure 10 shows the graph evaluating a change in the number of frequent patterns discovered as the number of data increases. The number of discovered frequent patterns increases similarly to the shape of $O(N)$, but it is not directly proportional. This is because a significant part contributes to the improvement of support for previously discovered patterns although the number of discovered patterns increases as the number of data increases.

### 5.3 Minimum spatial support count($\sigma$)

The minimum spatial support count is the value for determining which word is frequent in any cell. For example, if the minimum spatial support count ($\sigma$) is 2, cells where a word called 'nyc' appears in at least 2 social data become dense cells. Therefore, as the minimum spatial support count is smaller, the number of dense cells is larger, and as the minimum spatial support count is larger, the number of dense cells is smaller.

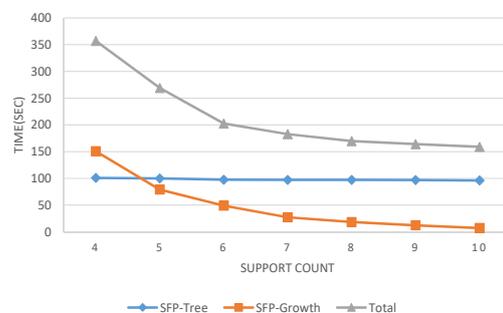

**Figure 11. Time of frequent wordset discovery by σ**

Figure 11 is the graph where the SFP-Tree construction time (+time to discover a dense cluster) according to minimum spatial support count (σ) increase is compared with SFP-growth execution time change. Here, the SFP-growth time is a scale of 1/1000. The SFP-Tree construction time includes the disk access time, and it is performed by O (N). Since N is fixed in this experiment, there is no change in the time according to the minimum spatial support count change. SFP-growth execution time decreases in the size of inverse O (NlogN) as shown in the figure, and this is because the number of recursive arithmetic operations decreases fast as the number of dense cells and dense clusters is smaller.

## 6. Conclusion

In this study, the SFP-Tree/SFP-Growth algorithms were proposed. These algorithms enabled spatial frequent pattern mining from social data and presented the vision for various research areas [19]. At first, the spatial transaction database is the best application target. For example, Walmart had stores in 4,135 areas in U.S. and 6,288 stores in other areas in 2013. At this time, mining of commodity transaction data which has code per branch store and relatively frequent items on a specific branch store is available. Also, it can play a role of base algorithm for finding the trend of regional public opinion or a regional event on the social network. While only global frequent patterns that are not spatially classified are discovered in normal social network, the proposed algorithms can apply to see in which trend spatially frequent patterns change over time. Also, it was proven through the experiments that the suggested algorithms carried out the pattern mining efficiently within a linear range according to data increase.

In future studies, it can be expanded to explore a spatial frequent pattern in disk-based system or big data cluster system. In this study, it was assumed that the SFP-tree based SFP-growth algorithm was executed in a limited memory resource of the computing system, but it is necessary to expand it to an algorithm which is executed for massive data based on the disk. Also, a study to expand the algorithm in parallel to enable the operation of algorithm in a cluster system, such as MapReduce of Hadoop, for larger social network is necessary. On the other hand, a study to explore continuous spatial frequent patterns based on sliding window for social data that has the form of continuous stream is also necessary.

**Acknowledgements** This research was supported by a Grant (14NSIP-B080144-01) from National Land Space Information Research Program funded by Ministry of Land, Infrastructure and Transport of Korean government.